\documentclass[sigconf,natbib=true,nonacm]{acmart}

\AtBeginDocument{%
  }

\setcopyright{acmlicensed}
\copyrightyear{2018}
\acmYear{2018}
\acmDOI{XXXXXXX.XXXXXXX}
\acmConference[IR 'XX]{Make sure to enter the correct
  conference title from your rights confirmation email}{June 03--05,
  2018}{Woodstock, NY}
\acmISBN{978-1-4503-XXXX-X/2018/06}

\usepackage{xspace}
\usepackage{subcaption}
\usepackage{booktabs}
\usepackage{multirow}
\usepackage{algorithm, algorithmic}
\usepackage{enumitem}

\begin{document}

\newcommand{\synapse}[0]{\textsc{Synapse}\xspace}

\title{\synapse: Evolving Job-Person Fit with Explainable Two-phase Retrieval and LLM-guided Genetic Resume Optimization}


\author{Ansel Kaplan Erol\footnotemark, Seohee Yoon, Keenan Hom, Xisheng Zhang}
\renewcommand{\shortauthors}{Erol et al.} 
\affiliation{%
  \institution{Georgia Institute of Technology}
  \city{Atlanta}
  \state{GA}
  \country{USA}
}

\renewcommand{\shortauthors}{Trovato et al.}


\begin{abstract}
Modern recruitment platforms operate under severe information imbalance: job seekers must search over massive, rapidly changing collections of postings, while employers are overwhelmed by high-volume, low-relevance applicant pools. Existing recruitment recommender systems typically rely on keyword matching or single-stage semantic retrieval, which struggle to capture fine-grained alignment between candidate experience and job requirements under real-world scale and cost constraints. We present \synapse, a multi-stage semantic recruitment system that separates high-recall candidate generation from high-precision semantic reranking, combining efficient dense retrieval using FAISS with an ensemble of contrastive learning and Large Language Model (LLM) reasoning. To improve transparency, \synapse incorporates a retrieval-augmented explanation layer that grounds recommendations in explicit evidence. Beyond retrieval, we introduce a novel evolutionary resume optimization framework that treats resume refinement as a black-box optimization problem. Using Differential Evolution with LLM-guided mutation operators, the system iteratively modifies candidate representations to improve alignment with screening objectives, without any labeled data.  Evaluation shows that the proposed ensemble improves $nDCG@10$ by 22\% over embedding-only retrieval baselines, while the evolutionary optimization loop consistently yields monotonic improvements in recommender scores, exceeding 60\% relative gain across evaluated profiles. We plan to release code and data upon publication.
\end{abstract}




\maketitle
\footnotetext{Corresponding author email: aerol3@gatech.edu}

\section{Introduction} 
The digitization of hiring has turned recruitment into a large-scale information retrieval problem, in which candidates and employers must navigate vast, evolving collections of semi-structured documents under tight time and cost constraints \cite{guo2017deeprec,liu2009learning}. Job seekers confront an unmanageable flood of postings across aggregators and company portals, while employers must sift through thousands of applications per role. The resulting volume erodes retrieval precision and creates inefficiencies on both sides of the market \cite{lops2011content,bevara2025resume2vec}. Automated application tools have intensified the problem, increasing applicant volume without improving relevance, overwhelming ranking systems and decision processes \cite{newsroom2025,mashayekhi2024survey}.
Despite this escalation in scale, most candidate-facing tools still rely on keyword search and simple filtering heuristics \cite{bevara2025resume2vec}. These approaches struggle with latent semantic alignment, as equivalent skills are expressed through inconsistent titles, shifting taxonomies, and domain-specific jargon, widening the ``vocabulary gap.''
Building higher-fidelity recommender systems for recruitment introduces additional structural challenges. Large-scale, publicly available relevance judgments linking candidates to postings are scarce, limiting the applicability of collaborative filtering and supervised learning-to-rank methods \cite{qin2024confit}. Second, the job corpus itself is highly dynamic, requiring continual indexing of millions of postings \cite{zhao2021scale}. Finally, while deep semantic models and LLMs improve matching quality, their computational costs make large-scale pairwise evaluation prohibitive for interactive systems, with per-document costs up to \$100 \cite{vaishampayan2025llmscoring}.
To address these constraints, we propose \synapse, a semantic recommender system designed specifically to match job seekers with relevant postings. \synapse utilizes a multi-stage architecture: dense vector representations first efficiently retrieve a subset of relevant jobs from the corpus, followed by a high-precision neural reranking stage using full context.
Unlike employer-centric filtering tools, \synapse prioritizes user agency. We employ Retrieval-Augmented Generation (RAG) to provide grounded, extractive explanations for ranking decisions, addressing trust issues in opaque neural rankers \cite{lewis2020rag, vaishampayan2025llmscoring, iso2025bias}. Additionally, we introduce a resume improvement tool inspired by evolutionary optimization, which uses language models as structured mutation operators, allowing applicants to iteratively refine their resumes to better align with the system's scoring function \cite{lehman2022evolution}.
The contributions of this work are:
\begin{enumerate}[nosep, leftmargin=*]
    \item A scalable, multi-stage semantic retrieval architecture for resume--job matching that integrates dense retrieval, contrastive reranking, and LLM-based reasoning.
    \item An explainable recruitment recommender that produces retrieval-grounded justifications for ranking decisions.
    \item A novel evolutionary resume optimization framework that leverages LLM-guided mutation operators to iteratively improve alignment with automated screening objectives.
\end{enumerate}
\vspace{-1em}

\section{Background and Related Work}
Automated resume--job matching is commonly studied as a specialized instance of document retrieval and recommendation, with heterogeneous, semi-structured text and limited relevance supervision. Early systems relied on rule-based filters and manual features, including keyword overlap and skill taxonomies \cite{lops2011content}. While effective for explicit constraints, such approaches were brittle to lexical variation and struggled to generalize across domains.

Advances in representation learning shifted the field toward latent semantic modeling of resumes and job postings. Neural encoders enabled semantic retrieval by capturing contextual similarity between candidate experience and job requirements \cite{belkin1992information}. More recent work frames resume--job matching as a relevance estimation problem aligned with learning-to-rank paradigms, where models score resume--posting pairs directly \cite{liu2009learning}.

\textbf{Embedding-based Retrieval.}
Embedding-based methods form the backbone of large-scale retrieval in recruitment systems. Models such as \texttt{Resume2Vec} encode resumes into dense vector representations, enabling approximate nearest-neighbor search that mitigates vocabulary mismatches \cite{bevara2025resume2vec}. By leveraging self-attention mechanisms, these models capture semantic similarity even when equivalent skills are expressed using divergent terminology \cite{devlin2018bert}.

Despite their scalability, embedding-only approaches often lack precision. Dense similarity metrics can fail to enforce ``hard'' constraints, such as required certifications or seniority, and may conflate semantically similar but incompatible roles \cite{bevara2025resume2vec}. This trade-off motivates \synapse's multi-stage retrieval architecture, where embeddings are used for high-recall retrieval followed by expressive reranking models, an established retrieval technique \cite{guo2017deeprec}.

To address poor precision in embeddings, recent frameworks leverage \textbf{contrastive learning} techniques, including data augmentation, hard-negative mining, and hypothetical embeddings to learn robust representations under sparse supervision \cite{qin2024confit, qin2025confitv2}. Production recruitment systems typically employ multi-stage cascaded pipelines to handle the scale of millions of postings while optimizing latency-quality trade-offs \cite{zhao2021scale,guo2017deeprec}. \synapse mirrors this tiered approach, integrating different types of reasoning in a scalable architecture.

\textbf{LLM-based Reranking and Explainability.}
Large Language Models (LLMs) have recently been explored as relevance estimators for resume--job matching. Operating over full document context, LLMs can perform nuanced reasoning, such as mapping project leadership experience to abstract managerial requirements \cite{vaishampayan2025llmscoring}. These capabilities make them attractive as rerankers in low-throughput settings. However, empirical studies highlight limitations that hinder direct deployment at scale. LLM-based scores can exhibit instability across candidates and reflect education and prestige biases \cite{iso2025bias}. Further, the computational cost of LLM evaluation is prohibitive for large corpora \cite{vaishampayan2025llmscoring}. Recent work therefore emphasizes grounding LLM outputs in retrieved evidence via Retrieval-Augmented Generation (RAG), improving transparency and reliability \cite{lewis2020rag}.


\section{The \synapse System}

    

\begin{figure}[t]
    \centering
    \includegraphics[width=0.95\columnwidth]{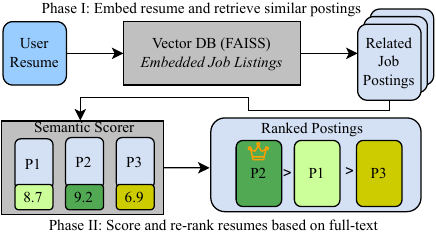}
    \caption{The \synapse recommender pipeline combines high-recall dense retrieval with precise semantic reranking.}
    \label{fig:recommender_pipeline}
\end{figure}

The \synapse system for recommending relevant job postings given an job seeker's resume is designed to balance semantic fidelity with scalability in person--job matching. It combines a multi-stage recommender pipeline with an iterative resume optimization utility, enabling both efficient retrieval over large corpora and fine-grained semantic alignment. The system operates over a corpus of approximately 120{,}000 job postings scraped from LinkedIn (see Sec. \ref{subsec:setup}), stored in a \texttt{SQLite} relational database for structured access and indexed using vector search for low-latency retrieval.

\subsection{Phase I: Candidate Retrieval}
The first stage of the \synapse pipeline performs broad, high-recall retrieval across the full job corpus. Resumes and job postings are encoded into a shared semantic space using a pretrained Sentence-BERT model (\texttt{all-MiniLM-L6-v2}) \cite{devlin2018bert}. Multiple attributes, including title, company, skills, and full description—are concatenated into a single document representation prior to encoding. Resumes are similarly fused from structured sections such as experience, education, and skills.

The resulting dense vectors are stored in a \texttt{FAISS} (Facebook AI Similarity Search) index which maps embeddings to posting ids, enabling approximate nearest-neighbor search for a given resume vector against a large posting corpus \cite{johnson2019faiss}. Given an uploaded resume, the system retrieves the top $K=25$ candidate postings based on cosine similarity. This stage prioritizes recall and throughput, aggressively reducing the search space while preserving relevant postings for downstream analysis.

\subsection{Phase II: Semantic Reranking}
The second stage performs high-precision reranking over the candidates retrieved in Phase I. Unlike shallow vector similarity, this stage leverages deeper semantic signals to better capture contextual alignment, including role seniority, domain specificity, and implicit qualification requirements. We explore several different approaches for reranking.

\textit{Approach 1: Pre-trained embedding similarity.} We embed documents similar to the original retrieval phase, but using a larger latent space for more fine-grained analysis.

\textit{Approach 2: Contrastive learning.} We train a self-supervised contrastive learning model using a \texttt{RoBERTa} backbone to generate dense vector representations. To generate training signal without labeled relevance data, we utilize a triplet loss framework where positive pairs were created via stochastic text augmentations—specifically random token masking, sentence shuffling, sentence dropout, and span deletion. Instead of relying on standard pooled embeddings, we employ a soft-alignment loss function that aggregates token-level similarities. This objective forces the model to maximize the alignment between a document and its augmented view while minimizing similarity to randomly sampled negative documents, thereby learning robust semantic features suitable for matching.

\textit{Approach 3: Language Models.} Next, we explore using language models to perform re-ranking. Traditional prompting that attempted to produce a numerical score for each resume-job pair was unsuccessful, even with rubrics and few-shot prompting. We achieve strong results when the LLM is instead prompted with a resume and two postings and asked to select the better aligned posting.

\textit{Ensembling.} On a small labeled validation set, we observed complementarity in the prior approaches and therefore ensembled them. We performed a sweep over various weighting and aggregation schemes, including by rank, z-score, min-max, softmax, Borda count, and the Reciprocal Rank Function (RRF).

\textit{By restricting this computation to a small candidate set, \synapse can afford more expressive models without incurring prohibitive cost. This two-stage design mirrors established multi-stage retrieval architectures, enabling a favorable trade-off between latency quality.}

\subsection{Retrieval-Augmented Explainability}
To improve transparency and provide users content for cover letters, \synapse incorporates a Retrieval-Augmented Generation (RAG) layer that produces grounded explanations for top recommendations. Relevant passages from both the resume and job description are retrieved and provided as context to an LLM. The model generates textual justifications that explicitly reference this evidence, explaining why a given position aligns with the candidate’s background. Constraining generation to retrieved content improves transparency and reduces hallucination risk. These explanations serve as user-facing feedback, highlights for cover letters, and as a diagnostic tool for understanding recommender behavior.

\subsection{Evolutionary Resume Refinement}

\begin{figure}[t]
    \centering
    \includegraphics[width=0.95\columnwidth]{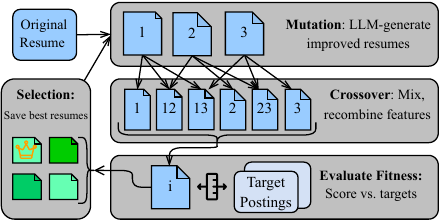}
    \caption{Resume improvement loop against target postings based on LLM-guided Differential Evolution.}
    \label{fig:evolutionary_loop}
\end{figure}

Beyond static matching, \synapse implements an iterative, evolutionary resume improvement tool. We frame resume refinement as a black-box optimization problem where a resume $r$ is an individual in a population $P$, and the match score against a set of target job postings $J$ serves as the fitness signal $f(r, J)$. The match score is computed as a normalized, weighted sum of (1) embedding distance (2) contrastive-learning document similarity, and (3) LLM-based pairwise ranking. Based on empirical tuning, we utilize a weighting scheme of $\mathbf{w} = [0.7, 0.15, 0.15]$, prioritizing alignment of latent embeddings while using higher-precision approaches as fine-grained discriminators.

\paragraph{Evolutionary Operators}
The system initializes a population of $N$ candidates via mutation of the user's base resume. At each generation $t$, we employ three operators:

\begin{itemize}
    \item \textbf{Mutation:} To balance exploration and exploitation, we implement an adaptive mutation schedule. The mutation operator rewrites resume sections with varying temperatures, ranging from synonym replacement and phrasing changes to structural rewriting. The probability of aggressive mutation increases in later generations to escape local optima.
    \item \textbf{Crossover:} Unlike traditional crossover, \synapse employs an LLM to semantically merge two parent resumes. The model is prompted to retain the strongest structural elements and keywords from both parents to generate a child entity.
    \item \textbf{Selection:} We select a new population of resumes, favoring higher-fitness resumes for the next generation.

\end{itemize}

At each generation, the top $k$ resumes remain unchanged, ensuring elitism and monotonic improvement of the fitness score.

\section{Evaluation}
\label{sec:evaluation}

We evaluate \synapse across three dimensions: computational efficiency, retrieval quality, and the effectiveness of our optimization strategies. We analyze the latency trade-offs between CPU and GPU implementations, assess ranking performance using Normalized Discounted Cumulative Gain (nDCG), and demonstrate the fitness improvements achieved through differential evolution.

\subsection{Experimental Setup}
\label{subsec:setup}

\paragraph{Datasets}
To simulate a realistic recruitment environment, we utilized two public datasets. For the job corpus, we ingested $120,000$ job postings collected from LinkedIn (2023-2024)\footnote{Source: \url{https://www.kaggle.com/datasets/arshkon/linkedin-job-postings/data}}, containing structured fields such as company, industry, title, and description. For candidate profiles, we utilized a diverse collection of $2,500$ resumes from LiveCareer\footnote{Source: \url{https://www.kaggle.com/datasets/snehaanbhawal/resume-dataset/data}}, spanning various experience levels and sectors.

\textit{Compute and Model APIs.} We tested our system on both CPU (Intel Xeon Gold 6226) and GPU (NVIDIA H100 HGX), using Gemini 2.0 Flash via the LiteLLM API for all LLM-based components.

\textit{Ground Truth \& Metrics}
Due to the lack of large-scale public datasets containing matched Resume-Job pairs with outcome labels, we constructed a small-scale, high-fidelity ground truth dataset. We collected application data from 10 distinct candidates over a 3-month period. Interactions were annotated with graded relevance scores: \textit{Offers} (+2), \textit{Interviews} (+1), and \textit{Rejections/Non-responses} (0).

To quantify ranking quality, we employ Normalized Discounted Cumulative Gain (nDCG@p), which evaluates the graded relevance of the top-$p$ retrieved items relative to an ideal ranking.

\subsection{Results}

\textbf{System Performance Evaluation.} We first evaluate the computational latency of our pipeline. Experiments were conducted on a workstation equipped with an Intel Xeon Gold 6226 CPU and an NVIDIA H100 HGX GPU. Table~\ref{tab:system_speed} details the execution time for both offline pre-computation and online inference phases. GPU acceleration provides significant gains in embedding and scoring, notably a massive reduction in Phase-II embedding scoring latency (from 0.57s to 0.026s, $\approx 22\times$). Despite parallelization, the end-to-end latency is dominated by LLM calls ($\approx 7$ seconds) via APIs.

\begin{table}[t]
\centering
\caption{System Performance Evaluation. Compares CPU (Intel Xeon Gold 6226) and GPU (NVIDIA H100 HGX).}
\label{tab:system_speed}
\resizebox{\columnwidth}{!}{%
\begin{tabular}{l c c}
\toprule
& \multicolumn{2}{c}{\textbf{Time ($\mu \pm \sigma$)}} \\
\cmidrule(lr){2-3}
\textbf{Task} & \textbf{CPU (Xeon)} & \textbf{GPU (H100)} \\
\midrule
\multicolumn{3}{l}{\textit{Offline / Pre-computation (minutes)}} \\
Embed 125k Postings & $94.06 \pm 3.19$ & $\mathbf{49.34 \pm 0.37}$ \\
Initialize FAISS & $0.51 \pm 0.01$ & $\mathbf{0.39 \pm 0.01}$ \\
\midrule
\multicolumn{3}{l}{\textit{Online / Inference (seconds)}} \\
{[I]} Embed Resume & $0.032 \pm 0.017$ & $\mathbf{0.006 \pm 0.002}$ \\
{[I]} Sim. Search ($k=20$) & $0.107 \pm 0.014$ & $\mathbf{0.065 \pm 0.005}$ \\
{[II]} Embed/Score & $0.570 \pm 0.025$ & $\mathbf{0.026 \pm 0.012}$ \\
{[RAG]} Gen. Explanation & $\mathbf{6.784 \pm 0.557}$ & $6.795 \pm 0.595$ \\
\midrule
\textbf{Full Pipeline} & $7.94 \pm .560$ & $\mathbf{7.03 \pm .595}$ \\
\bottomrule
\end{tabular}%
}
\end{table}
\smallskip
\noindent
\textbf{Ranking Quality Analysis} To assess the quality of our recommendations, we compared the baseline retrieval (Phase 1) against various reranking strategies and ensemble methods on our ground truth dataset. Table~\ref{tab:ndcg_results} reports nDCG@10 scores and  relative improvement ($\Delta$) over baseline. Individual reranking approaches performed moderately better than the pretrained embedding baseline, with 17.4\% improvement using contrastive learning. However, the weighted-average rank ensemble proved superior, yielding a 31.9\% improvement in nDCG@10. This suggests that while semantic embeddings capture general relevance, the combination of lexical matching, structural analysis, and LLM reasoning provides the most robust signal for candidate ranking.
\begin{table}[t]
\centering
\caption{Ranking Performance (nDCG) for k=10, k=20 postings, with relative improvement ($\Delta$) over embedding baseline.}
\label{tab:ndcg_results}
\resizebox{\columnwidth}{!}{%
\begin{tabular}{l c r c}
\toprule
\textbf{Method} & \textbf{nDCG@10} & \textbf{$\Delta$\% (@10)} & \textbf{nDCG@20} \\
\midrule
\multicolumn{4}{l}{\textit{Individual Approaches}} \\
Phase I Embed. Sim. & 0.5411 & --- & 0.5811 \\
Contrastive Learning & 0.6350 & +17.4\% & 0.6772 \\
LLM Ranks (Pairwise) & 0.5965 & +10.2\% & 0.6050 \\
\midrule
\multicolumn{4}{l}{\textit{Ensemble Approaches}} \\
\textbf{WAvg Rank (60/25/15)} & \textbf{0.7136} & \textbf{+31.9\%} & \textbf{0.7301} \\
HarmMean (75/15/10) & 0.6928 & +28.0\% & 0.7228 \\
WAvg MinMax (90/5/5) & 0.6342 & +17.2\% & 0.6842 \\
WAvg Z-Score (80/10/10) & 0.6142 & +13.5\% & 0.6342 \\
WAvg Softmax (70/15/15) & 0.5902 & +9.1\% & 0.6191 \\
Borda Count & 0.5859 & +8.3\% & 0.6063 \\
Reciprocal Ranking ($k=60$) & 0.5727 & +5.8\% & 0.5727 \\
\bottomrule
\end{tabular}%
}
\end{table}

Our evaluation highlights a critical trade-off between retrieval quality and system latency. While the Weighted Average Rank ensemble achieved the highest nDCG scores, it necessitates running multiple parallel scoring pipelines—including vector similarity, keyword matching, and LLM evaluation. Although GPU acceleration mitigates the cost of embedding-based scoring, the inclusion of LLM-based reranking introduces computational overhead that may be prohibitive for ultra-low-latency environments. Furthermore, a significant challenge in this domain is the scarcity of large-scale, labeled datasets for training and evaluation. Due to the proprietary nature of recruitment data, we utilized a small, manually annotated test set to validate our results; while necessary to establish a baseline, this constraint limits the statistical power of our findings compared to conventional supervised learning on massive datasets.

\begin{figure}[t]
    \centering
    \includegraphics[width=\linewidth]{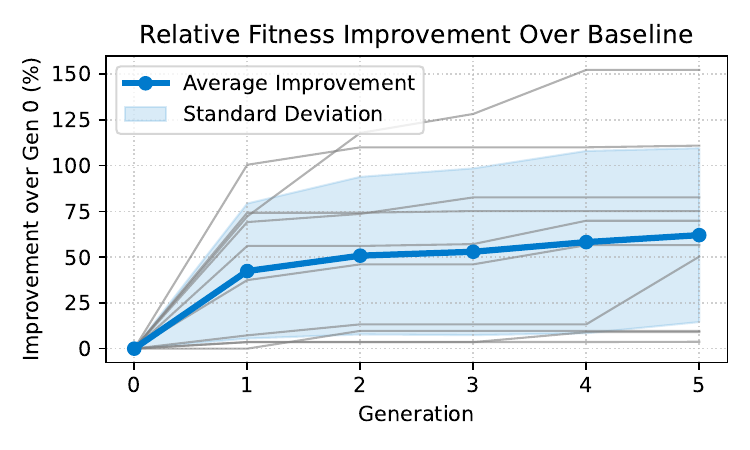}
    \caption{Relative fitness improvement across generations for ten resumes, the blue line representing the mean.}
    \label{fig:differential_evolution}
\end{figure}

\smallskip
\noindent
\textbf{Differential Evolution Optimization.} We evaluate our resume improvement tool across 100 random resumes and the top 5 postings generated for each resume. We configured the algorithm with a population of $N=8$ candidates, running for 5 generations with elitism count $k=2$, mutation rate of 0.7, and selection tournament size 3. As illustrated in Figure~\ref{fig:differential_evolution}, the optimization process resulted in monotonic improvements in fitness scores, and overall, achieved a median, mean, and upper-quartile improvement of 62\%, 68\%, and 92\% respectively.

\section{Conclusion}
This work introduced \synapse, an actionable and explainable framework for resume-job recommendation integrating (1) a two-phase recommendation system with vector retrieval and semantic reranking, (2) an explainability module, and (3) a novel differential evolution-inspired resume improvement pipeline. Our evaluations demonstrate that an ensemble of re-ranking techniques effectively overcomes data limitations to maximize retrieval quality, and that LLM-enabled evolutionary optimization successfully improves resumes against recommendation objectives. Future work can address data constraints with self-supervision and synthetic labels and explore other re-ranking strategies.

\bibliographystyle{ACM-Reference-Format}
\bibliography{sample-base}

\end{document}